\documentclass[iicol,sn-mathphys,Numbered]{sn-jnl}% Math and Physical Sciences Reference Style
\usepackage{manyfoot}%
\usepackage{graphicx}%
\usepackage{multirow}%
\usepackage{amsmath,amssymb,amsfonts}%
\usepackage{amsthm}%
\usepackage{xcolor}%

\definecolor{darkgreen}{rgb}{0,0.6,0.0}

\usepackage{anysize}
\newcommand\scalemath[2]{\scalebox{#1}{\mbox{\ensuremath{\displaystyle #2}}}}

\begin{document}

\title{Mixtures of self-propelled particles interacting with asymmetric obstacles}

\author[1]{\fnm{Mauricio} \sur{Rojas-Vega}}%\email{iauthor@gmail.com}
\author[2]{\fnm{Pablo} \sur{de Castro}}
\author[3]{\fnm{Rodrigo} \sur{Soto}}%\email{iauthor@gmail.com}

\affil[1]{\orgname{Institute of Science and Technology Austria}, \orgaddress{\street{Am Campus 1}, \postcode{3400}, \city{Klosterneuburg}, \country{Austria}}}

\affil[2]{\orgdiv{ICTP South American Institute for Fundamental Research \& Instituto de F\'isica Te\'orica}, \orgname{Universidade Estadual Paulista - UNESP}, \orgaddress{ \postcode{01140-070}, \city{S\~ao Paulo},  \country{Brazil}}}

\affil[3]{\orgdiv{Departamento de F\'isica, FCFM}, \orgname{Universidad de Chile}, \orgaddress{\street{Avenida Blanco Encalada 2008}, \city{Santiago}, \country{Chile}}}

\affil[]{\\ \textsuperscript{2}E-mail: pablo.castro@ictp-saifr.org (corresponding author)}

\abstract{
In the presence of an obstacle, active particles condensate into a surface ``wetting'' layer due to persistent motion. If the obstacle is asymmetric, a rectification current arises in addition to wetting. Asymmetric geometries are therefore commonly used to concentrate microorganisms like bacteria and sperms. However, most studies neglect the fact that biological active matter is diverse, composed of individuals with distinct self-propulsions.
Using simulations, we study a mixture of ``fast'' and ``slow'' active Brownian disks in two dimensions interacting with large half-disk obstacles.
With this prototypical obstacle geometry, we analyze how the stationary collective behavior depends on the degree of self-propulsion ``diversity'', defined as proportional to the difference between the self-propulsion speeds, while keeping the average self-propulsion speed fixed.
A wetting layer rich in fast particles arises. 
The rectification current is amplified by speed diversity due to a superlinear dependence of rectification on self-propulsion speed, which arises from cooperative effects.
Thus, the total rectification current cannot be obtained from an effective one-component active fluid with the same average self-propulsion speed, highlighting the importance of considering diversity in active matter. 
Finally, rectification alters particle evaporation and absorption by the layer, making the density of the dilute phase increase with the global density. Therefore, the steady state violates the lever rule, a result which is valid even for systems of identical particles.
}

\maketitle

\section{Introduction}
\label{intro}
Active particles---such as bacteria, tissue cells, and autophoretic colloids---spontaneously accumulate on the surfaces of obstacles even in the absence of attractive forces \cite{sepulveda2017wetting}. Similarly to motility-induced phase separation (MIPS) \cite{cates2015motility}, such \textit{active wetting} \cite{PhysRevE.107.014608} arises when active particles have a direction of motion that evolves stochastically but slowly, i.e., their direction of motion is \textit{persistent} \cite{wittmann2016active,de2021active,turci2021wetting, sepulveda2018universality}. For sufficiently large persistence times or densities, particles do not have time to find an escape route and thus become trapped between obstacles and other particles. Active wetting helps control surface adhesion and capillary properties of bacterial biofilms \cite{soto2014run,nie2021silico,fausti2021capillary}, whose formation makes bacterial colonies more resilient against antibiotics \cite{grobas2021swarming}.
\begin{figure*}
	\includegraphics[width=\textwidth]{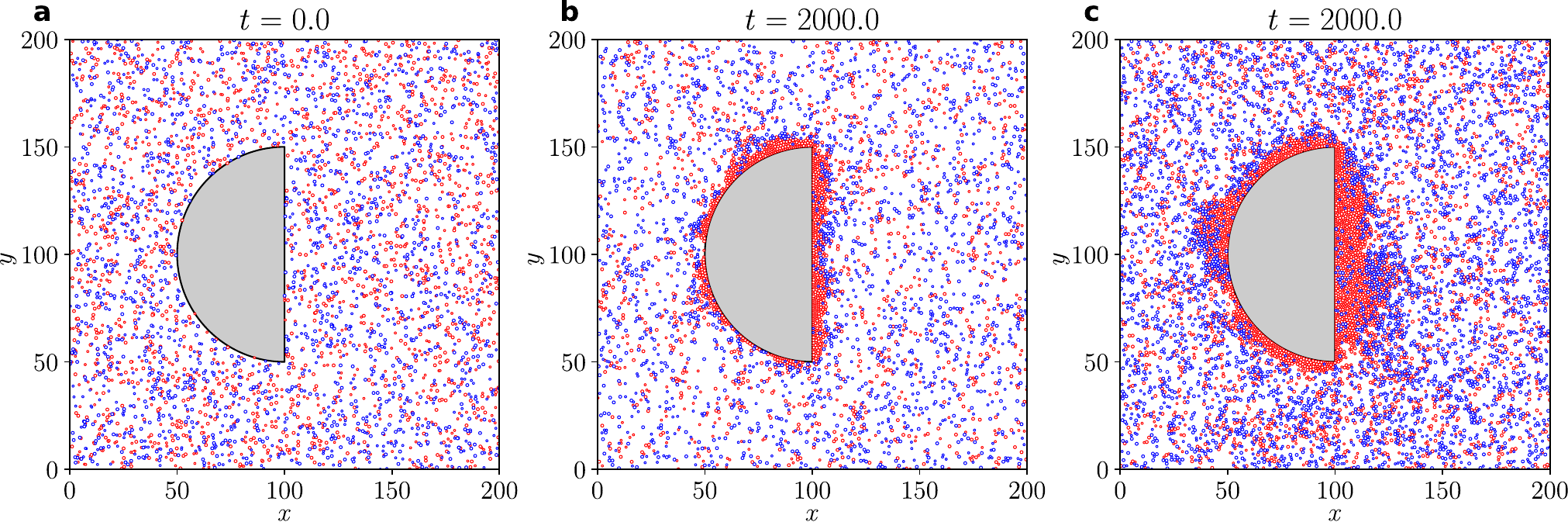}
	\caption{Snapshots of the system for speed diversity $\delta=0.8$ and $v_0=1$, with the faster particles in red and the slower ones in blue. The gray area corresponds to the obstacle. Periodic boundary conditions are used. (a) Homogeneous initial state for particle-occupied area fraction $\phi=0.13$. Configurations within the steady state, at $t=2000$, for (b) $\phi=0.13$ and (c) $\phi=0.26$. }
	\label{Snapshots}
\end{figure*}

In the case of \textit{asymmetric} obstacles, simulations and experiments show that active particles undergo directed motion \cite{reichhardt2017ratchet}, in addition to wetting. The spontaneous emergence of net particle transport due to environmental asymmetries, i.e., rectification currents, has constituted a central topic in both conceptual and technological contexts for decades \cite{reimann2002brownian}. More recently, research on rectification of self-propelled particles has gained momentum \cite{galajda2007wall,ghosh2013self,pototsky2013rectification,stenhammar2016light,zhu2020rectification,wagner2022steady,wagner2017steady,derivaux2022rectification,ai2016ratchet,savel2003controlling,savel2004manipulating,guidobaldi2014geometrical}. In Ref.~\cite{potiguar2014self}, an initially homogeneous collection of active Brownian particles in 2D was simulated in a regular array of half-disk rigid obstacles. Each obstacle was oriented along the same fixed direction as the others. The stationary average velocity, computed over all particles of the system, was found to be nonzero. Instead, an effective rectification current emerges since particles traveling from the curved to the flat side of the obstacle spend less time trapped by the obstacle than those in the opposite direction.

Ref.~\cite{potiguar2014self} considered identical active particles, i.e., particles with the same self-propulsion speed, rotational diffusion coefficient, and size. However, in natural colonies of bacteria and other microorganisms, a broad dispersion of motility parameters exists due to different ages, reproduction stages, shapes, sizes, and running modes \cite{ipina2019bacteria,berg2008coli,sparacino2020solitary,berdakin2013quantifying}. For either passive or active fluids, it is known that ``polydispersity'' or ``diversity'' of some particle attribute generates several new collective behaviors, including changing the nature and loci of phase diagram boundaries and introducing particle-type spatial segregation  \cite{PabloPeter1,PabloPeter2,PabloPeter3,decastro2019,stenhammar2015activity,agrawal2021alignment,kolb2020active,hoell2019multi,wittkowski2017nonequilibrium,takatori2015theory,grosberg2015nonequilibrium,curatolo2020cooperative,wang2020phase,van2020predicting,dolai2018phase,andrea2020,de2021active,kumar2021effect,pattanayak2020speed,semwal2021dynamics,singh2020phase,schmid2001wetting,brito2008segregation,costanzo2014motility,mones2015anomalous,mishra2012collective}. Still, it remains unclear what is the extent to which particle diversity affects rectification in active matter, as well as what are the mechanisms behind potential effects of particle diversity. A reason to study rectification in biologically-motivated active matter models is that, in complex biological environments, motile organisms commonly interact with obstacles of various shapes, as for example bacteria swimming around the internal structures of the host body where they live \cite{diluzio2005escherichia}.

In this work, we use two-dimensional (2D) simulations to investigate the collective behavior of a \textit{mixture}  of ``fast'' and ``slow'' active Brownian disks in the presence of a regular array of half-disk obstacles. In the stationary state, particles wet the obstacle and there is spatial segregation by particle type \cite{brito2008segregation,brito2009competition}. We observe the emergence of rectification currents together with the formation of vorticies near the obstacle corners \cite{pan2020vortex,caprini2020spontaneous}. We show that ``self-propulsion speed diversity'' (hereafter just \textit{speed diversity}) can have large effects on the collective particle dynamics. In our scenario, no external fields, hydrodynamic effects, or imposed alignment rules are present. Our goal is to understand how the mere existence of speed diversity may significantly alter collective observables such as rectification velocities and wetting or accumulation profiles. In particular, we focus on reporting and explaining why speed diversity amplifies rectification. The explanation for such effect is not straightforward since, while speed diversity is varied in our analysis, the system-average self-propulsion speed is kept unvaried. This leads us to the question of why rectification changes in the first place and why it becomes amplified by speed diversity. Crucially, our purpose here is not to seek an optimized rectifier but rather to identify the effects of speed diversity, when the other parameters remain fixed. For this reason, we will keep the geometry of the rectifier as simple as that of a half-disk.

This paper is organized as follows. In Section \ref{model} our model and simulation setup is laid out. In Section \ref{wetting}, we present our results for wetting and spatial segregation. In Section \ref{rectification}, we turn our attention to rectification effects and how they are connected with the appearance of circulating currents. Section \ref{Conc} brings our concluding remarks.

\section{Model and methods}
\label{model}
We consider $N$ active Brownian particles (ABPs) in 2D, each of them generically labelled by $i$. Half of the ABPs are ``fast'' particles, having self-propulsion speed ${v_i=v_\text{f}\equiv v_0 (1+\delta)}$. The other half are ``slow'' particles, with ${v_i=v_\text{s}\equiv v_0(1-\delta)}$. Therefore, the parameter $\delta \in [0,1]$ corresponds to the degree of speed diversity. Increasing $\delta$ makes the fast particles faster and the slow particles slower. If, say, a fast particle is momentarily at rest due to forces other than self-propulsion, it is still dubbed a fast particle. For ${\delta=0}$, all particles have identical self-propulsion speed and the system is said to be monodisperse or one-component. In the opposite limit, when ${\delta=1}$, the mixture becomes passive-active. For simplicity, global compositions other than $50$-$50\%$ are not considered, but generalization is straightforward. Therefore, on varying $\delta$, the system-average self-propulsion speed is kept at $v_0$, which is constant and independent of $\delta$. By doing so, the effects of speed diversity can be isolated. 
Artificial crystallization in 2D \cite{desmond2009random} is avoided by randomly assigning each particle one of two diameters, $d_\text{small}=d_0$ and $d_\text{large}=1.4 d_0$, uncorrelated with self-propulsion speeds. Since our focus is on the effects of speed diversity and we observed that size diversity effects were weak, the system is called just binary or bidisperse.

The position, $\boldsymbol{r}_i$, and orientation, $\theta_i$, of each ABP have a dynamics governed by the equations
\begin{equation}
\partial_t\boldsymbol{r}_i=v_{i}\,\hat{\boldsymbol{\nu}}_i+\mu \boldsymbol{F}_i,\quad \partial_t\theta_i=\eta_i (t), \label{eqmotion}
\end{equation}
where ${\hat{\boldsymbol{\nu}}_i=(\cos\theta_i, \sin\theta_i)}$ determines the self-propulsion force direction and $\mu$ is the particle mobility.  Also, $\boldsymbol{F}_i=\sum_{j\neq i} \boldsymbol{F}_{ij}+\boldsymbol{F}^{\text{obst}}_i$ is the net force on particle $i$ due to interactions with other particles and with the half-disk obstacle. The radius of the obstacle is $D/2$. The noise term $\eta_i (t)$ is Gaussian and white, with mean $\langle \eta_i(t)\rangle =0$ and correlation ${\langle \eta_i(t) \eta_j (t')\rangle =2\eta \delta_{ij}\delta(t-t')}$, where $\eta$ is the rotational diffusion coefficient.

The interparticle interactions are modeled by a soft repulsive Weeks-Chandler-Andersen (WCA)-like potential \cite{maloney2020clustering} defined in terms of the interparticle distance $r_{ij}$ as \footnote{The modified WCA potential used here has smooth second derivative, allowing it to be more suitable for (future) theoretical developments. Other repulsive potentials produce similar results.}
%\begin{equation}
%\scalemath{0.81}
%{
%U = \left\{
%\begin{array}{@{}l@{\thinspace}l}
%2^{\frac{3}{2}}\left(\cfrac{\sigma_{ij}}{r_{ij}}\right)^3-3\left(\cfrac{\sigma_{ij}}{r_{ij}}\right)^6+\left(\cfrac{\sigma_{ij}}{r_{ij}}\right)^{12}-\cfrac{3}{4}\,, &\quad r_{ij} \le  2^{\frac{1}{6}} \sigma_{ij},\\
%0, & \quad r_{ij}> 2^{\frac{1}{6}}\sigma_{ij} \\
%\end{array}
%\right.
%}
%\end{equation}
\begin{equation}
\scalemath{0.85}
{
U = \left\{
\begin{array}{@{}l@{\thinspace}l}
2^{\frac{3}{2}}\left(\cfrac{\sigma_{ij}}{r_{ij}}\right)^3-3\left(\cfrac{\sigma_{ij}}{r_{ij}}\right)^6+\left(\cfrac{\sigma_{ij}}{r_{ij}}\right)^{12}-\cfrac{3}{4}\,, &\quad r_{ij} \le  2^{\frac{1}{6}} \sigma_{ij},\\
0, & \quad r_{ij}> 2^{\frac{1}{6}}\sigma_{ij} \\
\end{array}
\right.
}
\end{equation}
with ${\sigma_{ij}\equiv\frac{1}{2}(d_i+d_j)}$, where $d_i$ is the diameter of particle $i$. For the particle-obstacle interaction between particle $i$ and the curved side, $d_j$ is replaced by $D$. For the interaction with the flat side, $d_j$ is replaced by zero.

We define two types of simulation analyses. For type I analysis, we keep $v_0=1$ and probe the effect of varying $\delta$. For type II analysis, we consider only one particle type and vary their speed $v_0$. Type II analysis will help make sense of the results coming out of type I analysis. For all particles, we use $\mu=1$, diameter scale $d_0=1$, $\eta=5\times 10^{-3}$, and the forward Euler method with time step $\Delta t=10^{-3}$. Initially, particles are distributed homogeneously at random positions with random velocity directions, independent of their type. The simulation is performed in a square box of side $L=200$ (such that $D=L/2$) and with periodic boundary conditions, simulating therefore an infinite regular array of identical obstacles. The obstacle is oriented along the $x$ axis and the centers of the flat side and of the simulation box coincide (see Fig.\ \ref{Snapshots}). 
The average free-particle persistence length is $\ell\equiv v_0/\eta$. For type I simulations, this gives $\ell=200$. For analyses of types I and II, the average free-particle persistence length is comparable to the system and obstacle sizes but still sufficiently small to avoid ballistic motion across the system in the presence of collisions.
The occupied area fraction $\phi$ is defined as the total area occupied by particles divided by the area of the simulation box minus the obstacle, i.e.,
\begin{equation} 
\phi=\frac{N}{2} \times\frac{\pi \left(d_\text{small}^2+d_\text{large}^2\right)/4}{L^2-\pi D^2/8}.
\end{equation}

The stationary behavior of any quantity of interest is obtained by averaging that quantity over distant time instants beyond $t=2000$. Stationary concentration fields for the total, slow, and fast particles are denoted respectively by ${\phi(\boldsymbol{r})}$, $\phi_\text{s}(\boldsymbol{r})$, and $\phi_\text{f}(\boldsymbol{r})$. These concentration fields are defined similarly to the area fraction $\phi$ but were calculated locally by using coarse-graining square boxes of side $2.5$. For this purpose, we counted the number of particle centers in each box, multiplied by the area of the corresponding particle, and then divided the result by the area of the coarse-graining box. For boxes that include a fraction of the obstacle, we calculated the available area using standard Monte Carlo integration. Concentration profiles were obtained by averaging the concentration fields over the direction parallel to each wall. For the curved side, the concentration profile is plotted against the radial distance to the wall, with all the radii used in the averaging meeting at the center of the flat side.

\section{Wetting and segregation}
\label{wetting}
We start with type I analyses where $v_0=1$ and $\delta$ is varied. For all parameter values, we observe the existence of a wetting layer of particles on the obstacle as well as a spatial segregation by particle type. In principle, these two particular phenomena are similar to those described in our previous work on fast and slow active particles interacting with a flat ``infinite'' wall \cite{PhysRevE.107.014608}. However, as we shall see, here the curvature and finite-size aspect of the obstacle walls play an important role in determining the wetting layer sizes, which in turn will affect the emerging rectification currents (absent for flat ``infinite'' walls) described in Section \ref{rectification}.

\begin{figure}[htb]
	\includegraphics[width=\columnwidth]{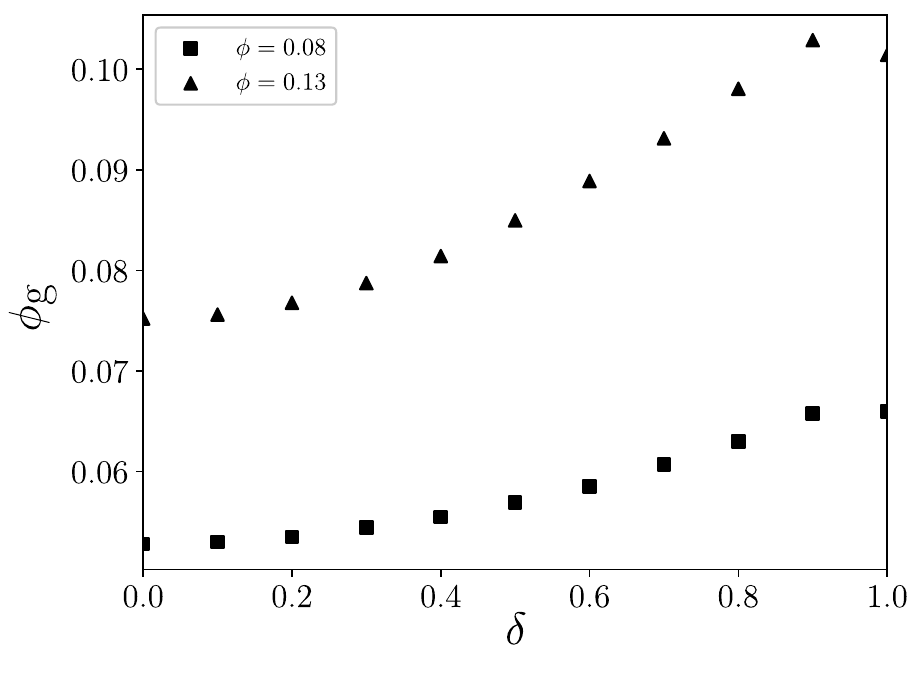}
	\caption{Particle area fraction in the gas phase $\phi_{\rm g}$ as a function of the diversity $\delta$ for two values of the global concentration $\phi$.}
	\label{fig.levelrule}
\end{figure}

Snapshots of the initial and stationary states are shown in Figs.~\ref{Snapshots}a and b for $\phi=0.13$ and $\delta=0.8$. Movie 1 of the Supplementary Material shows the transient dynamics, i.e., before the stationary state. Particles quickly accumulate around the obstacle forming an enveloping or wetting layer. The average thickness of the wetting layer stabilizes once the density of the ``gas''  (i.e., outside the layer) becomes sufficiently low that absorption and emission rates for the layer are equal. 
The lever rule of thermodynamics and thermodynamic-like systems, such as active systems under MIPS \cite{solon2018generalized}, states that the densities of the condensed and dilute phases are independent of the global density. Figure \ref{fig.levelrule} shows that the stationary particle area fraction in the gas phase, $\phi_{\rm g}$, depends on the global concentration $\phi$, revealing that the lever rule is not satisfied in this non-equilibrium system, not even in the steady state. 
%Figure \ref{fig.levelrule} shows that the stationary particle area fraction in the gas phase, $\phi_{\rm g}$, depends on the global  concentrations $\phi$. This is a signature that the well-known lever rule of thermodynamics and thermodynamic-like systems (such as active systems under MIPS \cite{solon2018generalized}), which states that the densities of the condensed and dilute phases are independent of the global density, is not satisfied in this non-equilibrium system, not even in the steady state. 
This is an effect of the stationary rectification current (see Section~\ref{rectification}) which makes the absorption and evaporation rates not thermal like. Instead, there is an active deposition and erosion of  particles by the rectification ``wind''. The violation of the lever rule is quantified as the quotient between the gas densities for two global densities, $\phi_{\rm g}(\phi=0.13)/\phi_{\rm g}(\phi=0.08)\approx 1.5$, with a weak increasing dependence with $\delta$.
%The violation of the lever rule, which appears here as a result of the rectification current, has been predicted to take place under quite general conditions~\cite{wittkowski2014scalar}.

Due to the different self-propulsion speeds, segregation emerges. The wetting layer increases with $\phi$, as shown in Fig.~\ref{Snapshots}c for $\phi=0.26$. In all cases studied below ($\phi=0.08$, $0.13$, and $0.26$, for various values of $\delta$), the stationary gas concentration is sufficiently low that no stationary clusters appear in the gas. That is, the residual gas density existent after the formation of the wetting layer is smaller than the necessary to produce MIPS \cite{cates2015motility}. Also, still during the transient regime, the condensation by heterogeneous nucleation responsible for wetting the obstacle is faster than an eventual MIPS. Hence, the only condensed phase in the system throughout the whole dynamical evolution is the wetting layer on the obstacle. This is also seen in Movies 2 and 3 of the Supplementary Material, which show, respectively, the transient and the steady state dynamics for $\phi=0.26$ with $\delta=0.8$. This lack of clusters in the bulk allows us to focus on investigating obstacle effects without getting obscured by additional aggregation effects.

\begin{figure}[!h]
	\includegraphics[width=0.95\columnwidth]{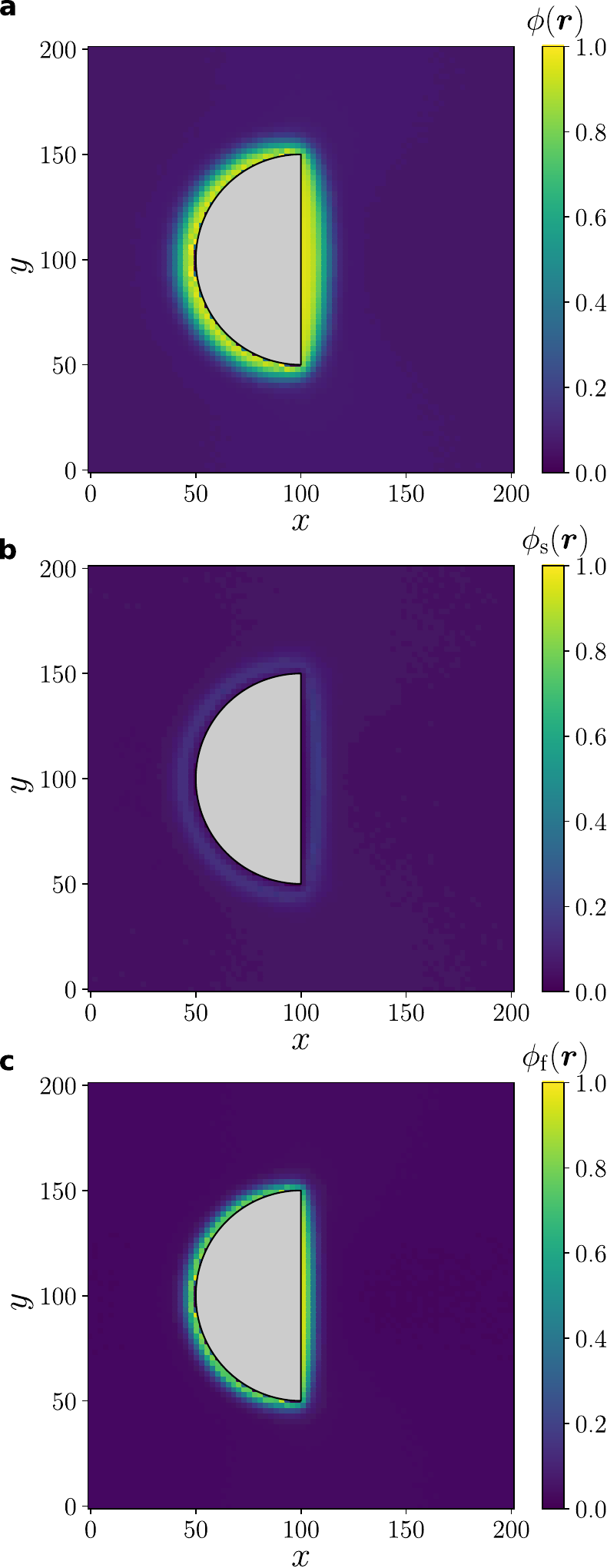}
	\caption{Stationary concentration fields for $\phi=0.13$ in a type I analysis. (a) Monodisperse system ($\delta=0$). (b) Slow and (c) fast particles for mixture case with $\delta=0.8$. }
	\label{ConcentrationFields}
\end{figure}

\begin{figure}
	\includegraphics[width=0.95\columnwidth]{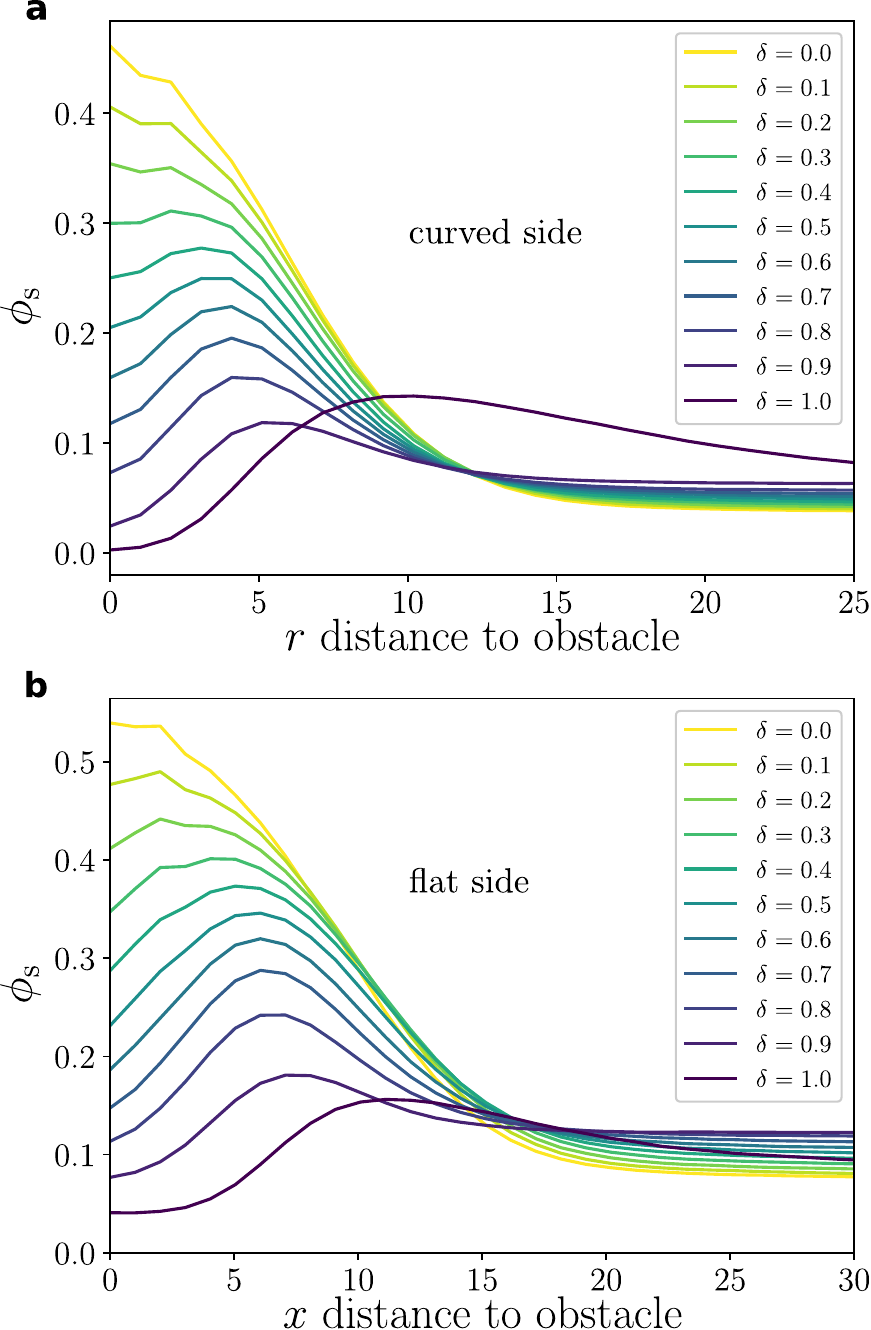}
	\caption{Stationary concentration profiles of slow particles for various values of $\delta$ (type I analysis) and $\phi=0.13$. (a) Curved and (b) flat side. }
	\label{Profiles}
\end{figure}

Fig.~\ref{ConcentrationFields} shows stationary concentration fields for a monodisperse systems and for a mixture. For $\delta>0$, we observe that $\phi_\text{s}(\boldsymbol{r})$ and $\phi_\text{f}(\boldsymbol{r})$ are significantly different from each other. The faster particles dominate the occupation closer to the obstacle, whereas the slower particles accumulate less sharply. Such spatial segregation depends on speed diversity as follows.
For the faster particles, the concentration decays monotonically towards the gas on both the curved and flat sides for any value of $\delta$ (concentration profiles not shown). 
For the slower particles, however, Fig.~\ref{Profiles} shows that the concentration profile develops a peak located away from the obstacle wall when $\delta\geq0.4$. This transition occurs when the slower particles become sufficiently slow that they can only accumulate on the boundary of the layer of faster particles instead of closer to the obstacle wall, into which they are too weak to penetrate. The peak is less pronounced on the curved side. This segregation picture will be an important ingredient for our main goal of understanding how rectification is affected by speed diversity. 

\begin{figure}
	\includegraphics[width=0.95\columnwidth]{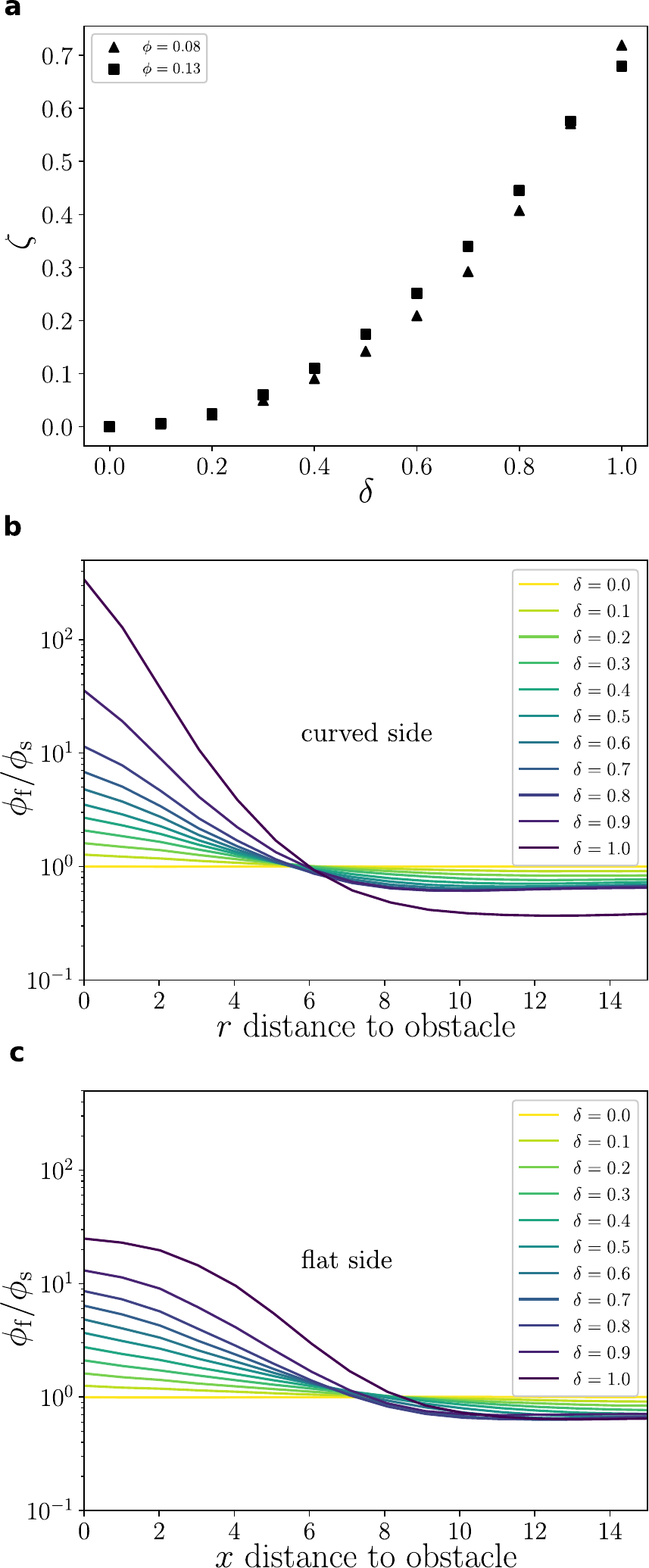}
	\caption{(a) Global segregation parameter for $\phi=0.08$ and $\phi=0.13$ versus $\delta$ (type I analysis). (b) Curved and (c) flat side concentration profiles ratios (fast over slow) for $\phi=0.13$ and various values of $\delta$.}
	\label{ratio}
\end{figure}

In order to further characterize how spatial segregation depends on speed diversity, Fig.~\ref{ratio}a shows the segregation parameter, a measure of the degree of segregation defined as \cite{brito2009competition}
\begin{equation}
\zeta= 1-\cfrac{\int \,\phi_\text{s}(\boldsymbol{r}) \phi_\text{f}(\boldsymbol{r}) \,dx\, dy }{\sqrt{\int \,\phi_\text{s}^2(\boldsymbol{r}) \,dx\,dy\int\, \phi_\text{f}^2(\boldsymbol{r}) \,dx\,dy}},
\end{equation}
and which takes into account the overlapping between the concentration fields. As such, $\zeta=1$ implies complete segregation and $\zeta=0$ means complete mixing, which in turn occurs only if $\phi_\text{s}(\boldsymbol{r})\propto \phi_\text{f}(\boldsymbol{r})$ everywhere. As shown in Fig.~\ref{ratio}a, complete segregation is never obtained, not even in the passive-active limit, since the gas has always a mix of both particle types due to particles escaping from the wetting agglomerate. 
%Fig.~\ref{ratio}a also shows that the degree of segregation in the low global concentration limit is almost independent of $\phi$. 

Segregation can also be quantified by the ratio between the slow and fast particles concentration profiles, which is shown in Figs.~\ref{ratio}b and c for the curved and flat side, respectively. For $\delta=1$, there are almost no passive particles near the wall and the curved side concentration ratio is more than one order of magnitude bigger than for $\delta=0.9$. This reveals that the case $\delta=0.9$ is not as close to a passive-active mixture as one might expect. In fact, Fig.~\ref{Profiles} already showed a significant difference in the behavior of the concentration profiles between $\delta=0.9$ and $\delta=1.0$. This can be understood by noticing that for $\delta=0.9$ the slow particles persistence length $v_\text{s}/\eta=v_0(1-\delta)/\eta=20$ is still comparable to other relevant length scales such as the wetting layer thickness, the obstacle size, and the system size, and therefore the slow particles in this scenario cannot be regarded as almost passive.

Generally, we observe that there is more segregation on the curved side than on the flat side because particles are less capable of penetrating and settling inside the curved-side layer. In fact, a particle with weak self-propulsion cannot easily remain at the curved-side wetting interface without being wiped out into the gas by the rectification ``wind'' (see Section~\ref{rectification} below); conversely, particles on the flat side can accumulate closer to the wall since the particle current on the flat side near the interface is not sufficient to wipe them out. This difference between the two sides in the wetting properties of the slow particles can be confirmed by looking once again at the concentration profile of slow particles in Fig.~\ref{Profiles} as $\delta=1$ is approached, where the concentration near the wall is practically zero on the curved side but not so on the flat side.

\begin{figure}
	\includegraphics[width=0.95\columnwidth]{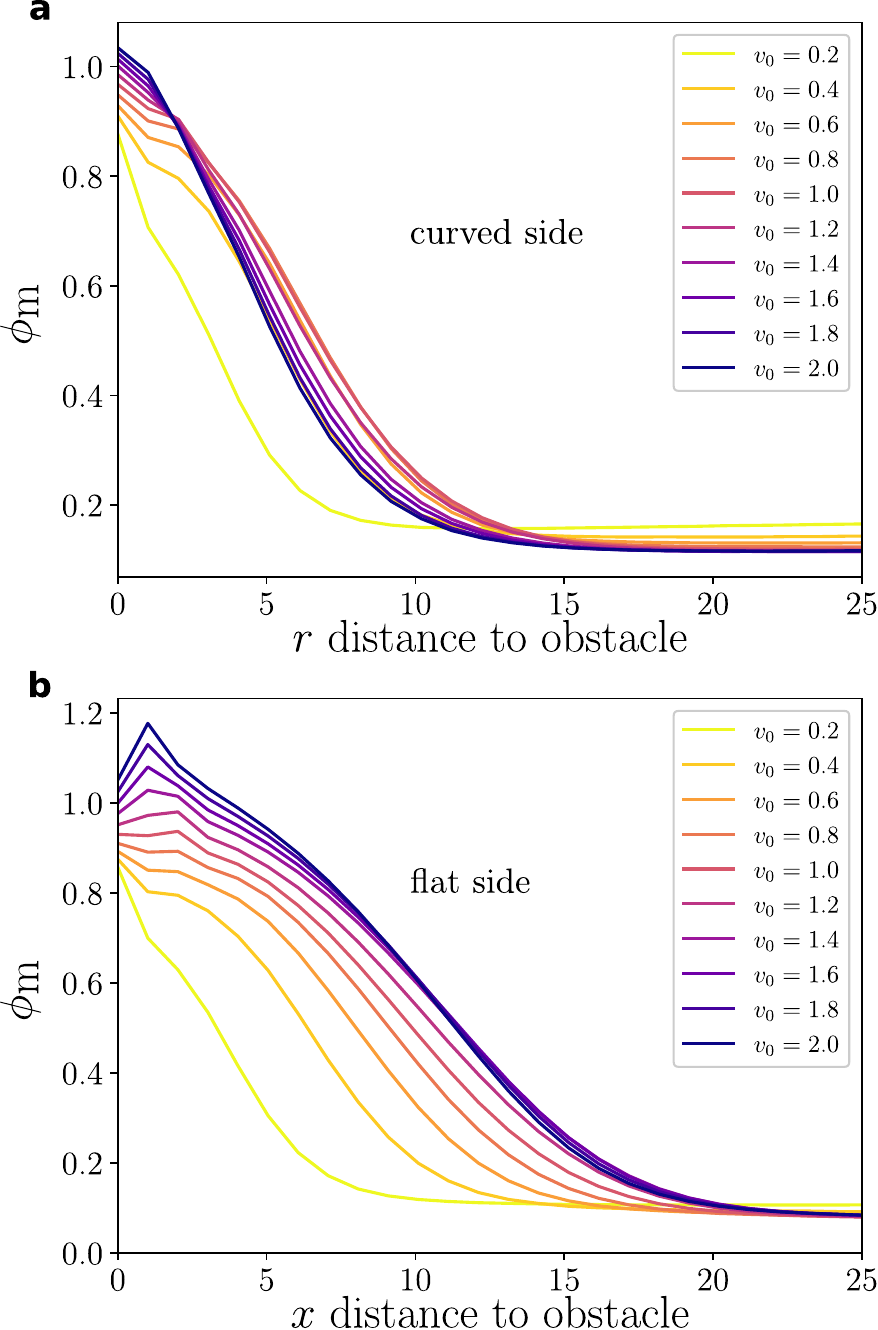}
	\caption{Stationary concentration profile, $\phi_{\rm m}$ (monodisperse case), for $\phi=0.13$, with all particles having equal self-propulsion speed $v_0$, which in turn is varied between the distinct curves. This further corroborates the latter observations (type II analysis). (a) Curved and (b) flat side.}
	\label{ProfilesTypeII}
\end{figure}

Finally, we notice that an analysis of type II further corroborates the latter observations In this type of analysis, all particles have equal self-propulsion speed $v_0$, which we vary. Figure~\ref{ProfilesTypeII} shows the concentration $\phi_{\rm m}$ versus distance to the wall for both obstacle sides. While for the flat side the wetting layer always increases with self-propulsion speed $v_0$, for the curved side the size of the layer depends non-monotonically on the self-propulsion speed $v_0$. This non-monotonicity is understood as follows. By increasing $v_0$, one increases both the particle's individuals tendency to wet as well as the rectification ``wind'' that wipes particles out of the wetting layer. For a certain value of $v_0$, the latter effect wins against the former. As a result, the size of the wetting layer starts to decrease with $v_0$. As we shall see, this observation helps make sense of the mechanisms involved in how the rectification velocity depends on speed diversity.

\section{Rectification and vorticity}
\label{rectification}

\begin{figure}[!h]
	\centering
	\includegraphics[width=\columnwidth]{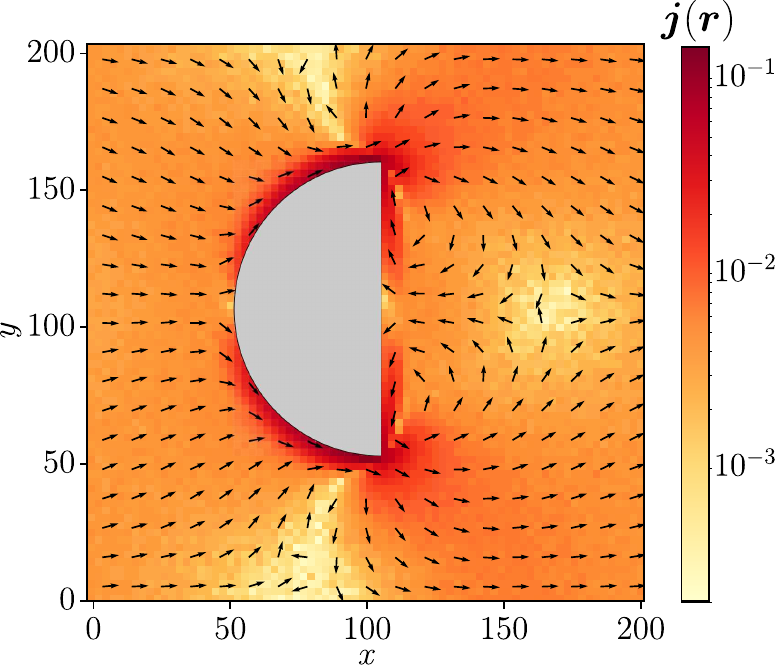}
	\caption{Total stationary current field $\boldsymbol{j}$ for $v_0=1$, $\phi=0.13$ and $\delta=0.8$. For clarity, the arrows have all the same size and only indicate the direction of the current, while the magnitude of $\boldsymbol{j}$ is given via the color legend.} 
	\label{CurrentField}
\end{figure}

The asymmetric shape of the obstacle implies that particles traveling to the right from the curved side will be trapped by the obstacle for less time than those traveling to the left from the flat side. Consequently, a rectification current is present in the stationary state as observed for the monodisperse case ($\delta=0$) \cite{potiguar2014self}. Here, we focus on rectification for active mixtures ($\delta>0$). Figure~\ref{CurrentField} shows the total stationary current (vector) field $\boldsymbol{j}(\boldsymbol{r})\equiv \phi(\boldsymbol{r})\,\boldsymbol{v}(\boldsymbol{r})$, where $\boldsymbol{v}(\boldsymbol{r})$ is the actual velocity field, not the self-propulsion velocity field. 
Global rectification along the $+x$ direction is indicated by the fact that most current arrows point to the right or have a large $+x$ component. Particles moving towards the right will slide on the curved side and are subject to a higher current flow in the wetting layer than in the gas. The highest local current is observed near the corners. As we approach the obstacle from the flat side, the local current changes direction (from $+x$ to $-x$) and, near the obstacle, becomes constrained to the $y$-axis only. Far from the obstacle, the current field changes to the $+x$ direction again.

\begin{figure}[!h]
	\includegraphics[width=0.95\columnwidth]{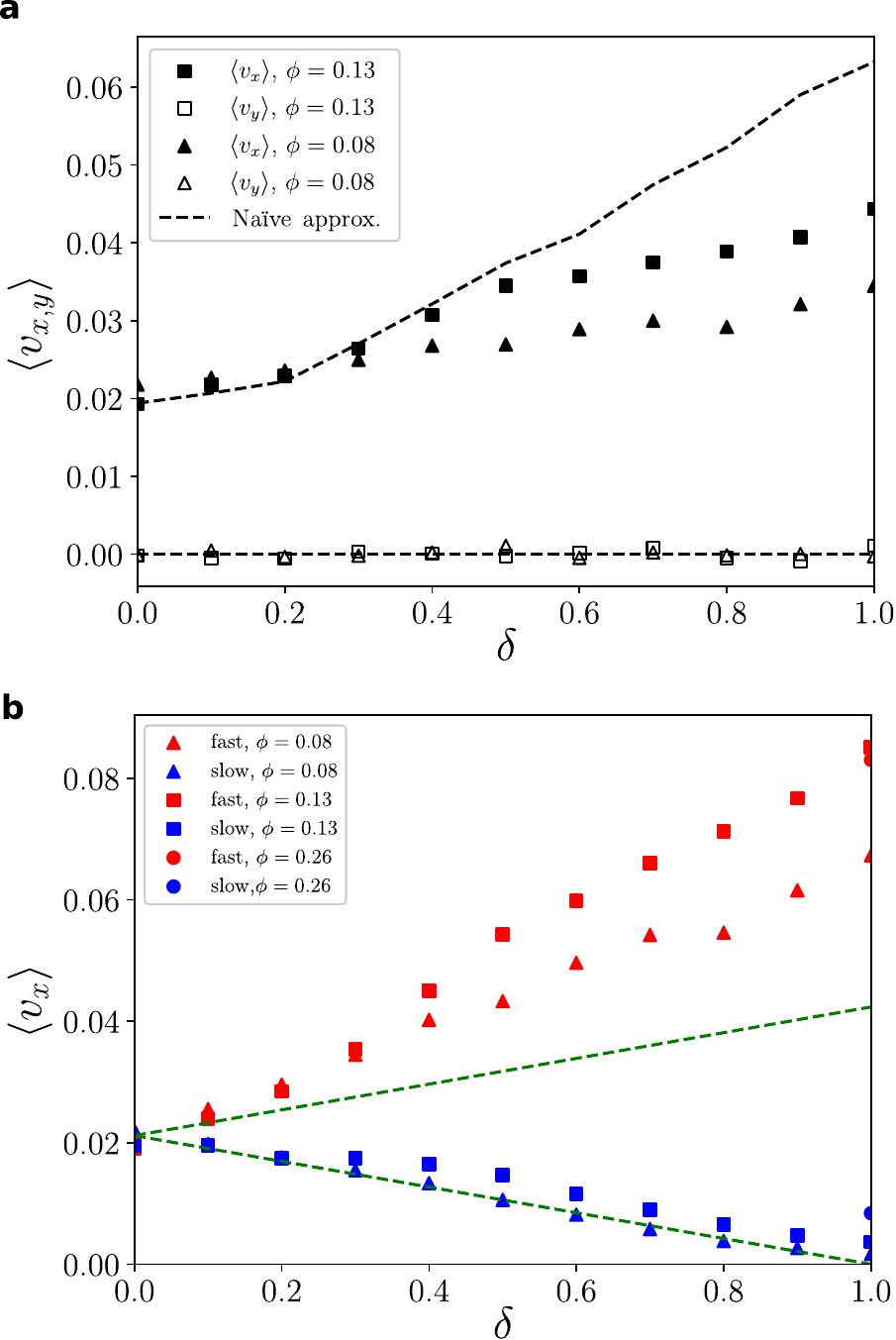}
	\caption{Stationary mean (actual) velocity averaged over particles and realizations as a function of $\delta$ (type I analysis) for $\phi=0.08$ and $0.13$.  (a) Total mean velocity in $x$ and in $y$. The dashed line is the na\"ive estimate for a mixture obtained by averaging the simulation results for monodisperse systems (type II analysis) with self-propulsion speeds $v_0(1+\delta)$ and $v_0(1-\delta)$, each contributing with $50\%$. (b) Mean velocity in $x$ for slow and for fast particles. The case $\phi=0.26$ has been included only for $\delta=1$, avoiding an overcrowding of the figure.  The dashed lines present the na\"ive dependence for the mean velocity of the fast and slow particles.}
	\label{MeanV}
\end{figure}

To investigate how rectification is affected by speed diversity, we show in Fig.~\ref{MeanV}a the mean velocity in $x$, $\langle v_x \rangle$, averaged over particles and over time instants within the steady state, as a function of $\delta$ (type I analysis). As a control, we also show that $\langle v_y \rangle$ is essentially zero, as expected. Notably, $\langle v_x \rangle$ increases with $\delta$, indicating an \textit{amplification} of rectification. Such amplification is induced solely by speed diversity since each particle type corresponds to $50$\% of all particles and thus the system-average self-propulsion speed does not change with $\delta$. By looking at $\langle v_x \rangle$ for each particle type in Fig.~\ref{MeanV}b, we see that, indeed, as $\delta$ increases, the fast particles undergo a rectification \textit{increase} which is larger than the rectification \textit{decrease} of the slow particles, even though their self-propulsion speeds were varied by the same magnitude $\delta$. Indeed, $\langle v_x \rangle$ for the fast particles is larger than the na\"ive dependence $1+\delta$, a manifestation of significant cooperative effects (more below). While for the slow particles at low $\phi$ the na\"ive dependence $1-\delta$ is followed, interactions dominate for higher $\phi$. As a result, the $\langle v_x \rangle$ of the slow particles is also larger than its na\"ive estimate $1-\delta$. 

\begin{figure}[!h]
	\includegraphics[width=\columnwidth]{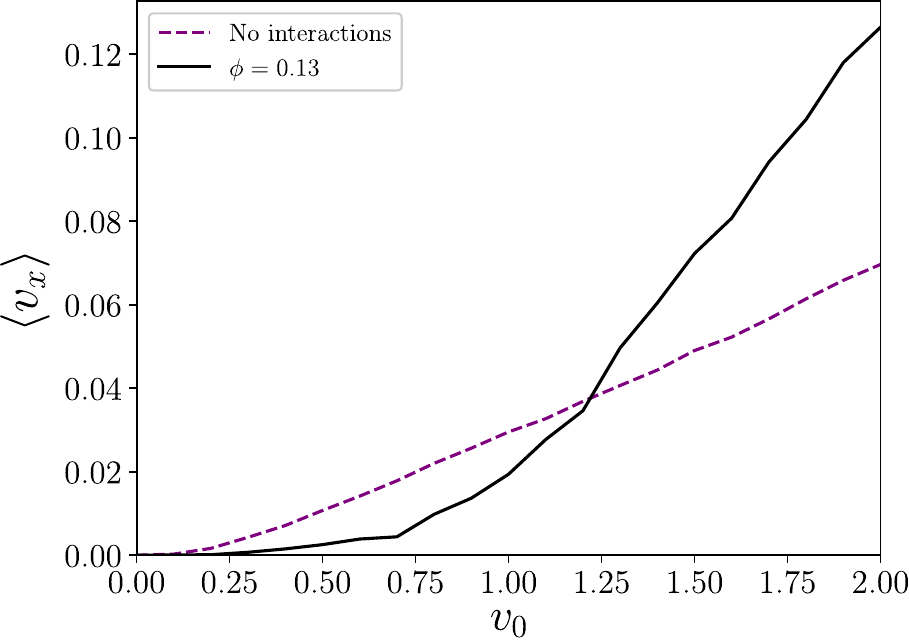}
	\caption{Stationary mean (actual) velocity as a function of $v_0$ (type II analysis). The dashed line is for the case with no particle-particle interactions and the solid line is for $\phi=0.13$.}
	\label{MeanVnew}
\end{figure}

To understand such rectification amplification induced by speed diversity, let us first take a few steps backs. Fig.~\ref{MeanVnew} revisits the monodisperse scenario, once again, in order to better understand how cooperative effects on rectification arise and depend on the self-propulsion speed $v_0$ (type II analysis). First, we measure the rectification velocity $\langle v_x \rangle$ in the absence of particle-particle interactions. In this case, each particle is independent of each other and therefore the wetting layer is virtually zero. Still, rectification arises because a single particle spends more time blocked by the flat side than by the curved side. For intermediate and high $v_0$, where rotational diffusion plays a minor role, rectification increases linearly with $v_0$. In this scenario, the particle escapes the obstacle typically by ``coasting'' the obstacle wall until its end. As a result, $v_0$ alone controls the escape time. For the flat wall, particles escape along the $y$-axis, therefore without affecting rectification. For the curved side, however, particles escape along the $x$-axis with speed proportional to $v_0$, which explains the linear behavior of $\langle v_x \rangle$ as a function of $v_0$. For small $v_0$, rotational diffusion becomes important and particles become likely to leave the obstacle before reaching its corner. For low $v_0$, therefore, the linear regime of $\langle v_x \rangle$ versus $v_0$ disappears.

Extending type II analysis, we now turn on particle-particle interactions. Two scenarios are possible for the behavior of $\langle v_x \rangle$ versus $v_0$. 
For small $v_0$, the rotational diffusivity is significant, implying that the particles can momentarily block each other when changing directions by rotational diffusion.  This leads to $\langle v_x \rangle$ smaller than in the case without interactions. 
However, for higher $v_0$, rectification inevitably becomes strong. In this scenario, a particle that ceases from moving in the ``easy'' direction ($+x$) will eventually be forced back into that direction by the flow of surrounding particles moving in the easy direction. In other words, if a particle starts to move more along the $y$-axis, e.g., due to rotational diffusion, it will be dragged along the $x$-axis by strongly rectified particles. Such cooperative effect makes particles become even more rectified upon increasing $v_0$ than they would without particle-particle interactions. Besides that, as showed in Fig.~\ref{ProfilesTypeII}, there is a rectification ``wind'' which cleans the curved side more than the flat side. For high $v_0$, the number of trapped particles increases with $v_0$ for the flat side and decreases for the curved side. This behavior further contributes to rectification since the flat side traps for longer. As a result of these two cooperative mechanisms, the rectification velocity $\langle v_x \rangle$ increases superlinearly with $v_0$, that is, more strongly than linearly. In summary, for low $v_0$, one has that $\langle v_x \rangle$ is smaller than it would be in the ideal gas scenario. For high $v_0$, $\langle v_x \rangle$ is bigger than for the ideal gas.

A na\"ive approach to try to explain the behavior of $\langle v_x \rangle$ as a function of $\delta$ (type I analysis, Fig.~\ref{MeanV}) is the following. We fix $v_0=1$ and calculate the average between $\langle v_x \rangle$ for a \textit{monodisperse} system with self-propulsion speed $v_0 (1+\delta)$ and $\langle v_x \rangle$ for a monodisperse system with self-propulsion speed $v_0(1-\delta)$, each contributing with $50\%$, using the simulation data in Fig.~\ref{MeanVnew}. The result is shown in Fig.~\ref{MeanV}a (dashed line). Such wrong averaging procedure neglects fast-slow interactions. As a consequence, $\langle v_x \rangle$ lies above the measured $\langle v_x\rangle$ of the actual mixture system. Still, the increase of $\langle v_x \rangle$ with $\delta$ seen in Fig.~\ref{MeanV}a is captured. The reason why the na\"ivelly calculated $\langle v_x \rangle$ is higher than in the case with fast-slow interactions is because, due to spatial segregation, more fast particles get trapped in the layers than slow ones. Once trapped, those particles become temporarily unavailable to contribute to the rectification current.

Furthermore, Fig.~\ref{MeanV}b shows that $\langle v_x \rangle$ for the slow particles does not vanish completely at the active-passive limit $\delta=1$. In fact, the passive particles continue to contribute positively to the total $\langle v_x \rangle$. This kind of behavior where the motion of passive particles is enhanced by active ones has been previously reported in the context of MIPS: the presence of active particles in fact induces \textit{clustering} for the passive ones \cite{stenhammar2015activity,kolb2020active,gokhale2022dynamic}. For infinite flat walls, we showed in Ref.\  \cite{PhysRevE.107.014608} that, during the transient dynamics before the stationary state, active particles can induced a finite degree of \textit{wetting} for passive particles. Here, we show that active particles induce a finite degree of \textit{rectification} for passive ones. Such ``passive rectification'' increases with $\phi$, as shown in Fig.~\ref{MeanV}b by varying $\phi$ for $\delta=1$. 
Also, notice that $\langle v_x \rangle$ for the fast particles increases more than just proportionally to $v_0 (1+\delta)$, in agreement with the cooperative effects presented in Fig.~\ref{MeanVnew}.

\begin{figure}[h!]
	\centering
		\includegraphics[width=\columnwidth]{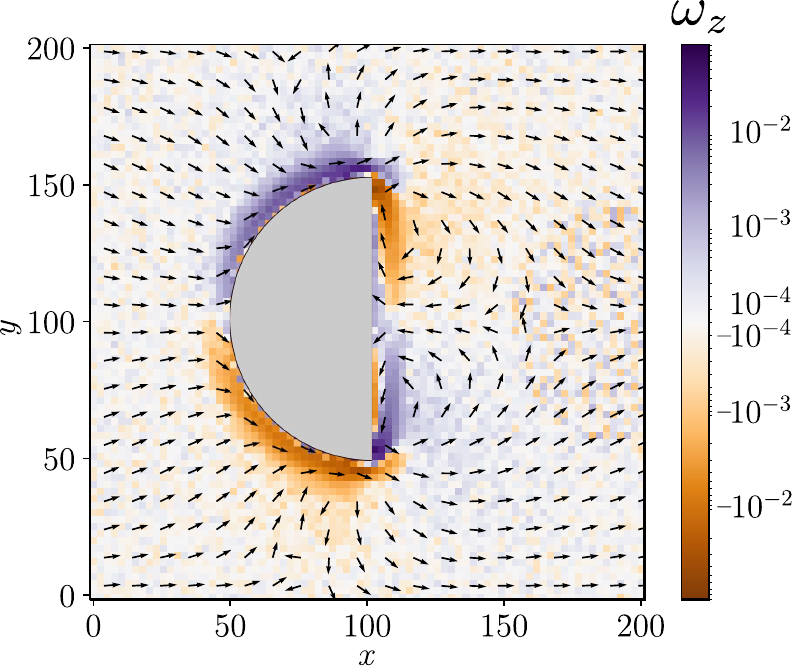}
	\caption{Vorticity-like field defined by the $z$-component of $\boldsymbol{\omega}( \boldsymbol{r} )\equiv\nabla \times \boldsymbol{j}(\boldsymbol{r})$, i.e., $\omega_z$, for ${\delta=0.8}$, $\phi=0.13$ and $v_0=1$, with positive (negative) values meaning counterclockwise (clockwise) rotation. The color scheme is shown on the right in symmetric logarithmic scale. The arrows, as in Fig.~\ref{CurrentField}, show the direction of the total particle current field.}
	\label{vorticityfieldfig}
\end{figure}

\begin{figure}
	\centering
	\includegraphics[width=0.95\columnwidth]{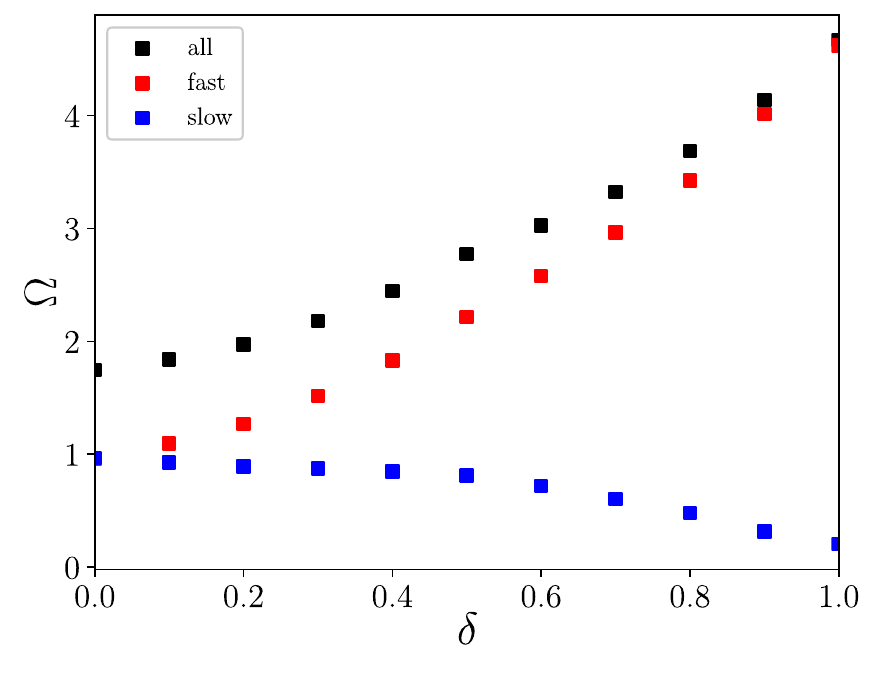}
	\caption{Total global magnitude of vorticity-like field as a function of $\delta$ (type I analysis) for $\phi=0.13$ and $v_0=1$. The partial vorticities, for fast and slow particles, are obtained from the partial current fields $\boldsymbol{j}_\text{f/s}(\boldsymbol{r})\equiv \phi_\text{f/s}(\boldsymbol{r})\,\boldsymbol{v}_\text{f/s}(\boldsymbol{r})$.}
	\label{TotalVorticity}
\end{figure}

Revisiting the spatial field of particle current for the active-active mixture case ($0<\delta<1$), we notice on the flat side the existence of a nonvanishing total local current moving away from the obstacle center along the $y$-axis. This behavior is connected to the appearance of a pair of vortices around each obstacle corner. For each pair, a vortex is produced by a ``rectification jet'' of particles that slide on the curved side while the other vortex is formed by particles that slide on the flat side. The latter particles move away from the obstacle center, along the $y$-axis. For each side, once the obstacle wall ends, the particles start to interact directly with those that were sliding on the other side traveling at a right angle. As a consequence, velocities become reoriented, thus generating the vortices. The existence of such stationary vortices is confirmed by defining the vorticity-like field
\begin{equation}
\boldsymbol{\omega}( \boldsymbol{r} )\equiv\nabla \times \boldsymbol{j}(\boldsymbol{r})
\end{equation}
and plotting its $z$ component in Fig.~\ref{vorticityfieldfig}. For convenience, this is not exactly the vorticity since it is the curl of the current field, not of the velocity field. Since the rectification velocity itself increases with $\delta$, the degree of vorticity is expected to increase with $\delta$ as well since the vortices depend on the existence of the rectification jets. Fig.~\ref{TotalVorticity} shows the magnitude of the global vorticity, defined as $ \Omega \equiv \int \omega_z (\boldsymbol {r}) \, dx\, dy $, and the contributions coming from the slow and fast particles, where we keep the signs in $\omega_z$. This confirms that the intensity of the total vorticity is also amplified by speed diversity. 

\section{Conclusions}
\label{Conc}
Here we considered an active mixture of fast and slow active particles in the presence of asymmetric obstacles, with a curved and a flat side. We identified the spontaneous emergence of wetting layers, spatial segregation, violation of the lever rule, rectification currents, and vortices. As such, this problem arises as an interesting playground for studying several phenomena at once. Using simulations, we showed how the degree of diversity of self-propulsion speeds alters these phenomena, both quantitatively and qualitatively. In particular, we found that, above a certain degree of speed diversity, a peak in the concentration profiles of the slower particles arises far from the wall, on top of the faster particles layer. This is because self-propulsion becomes too weak to allow for accumulation near the wall. Due to a rectification ``wind'', the curved side can be partially wiped out. As a result, the size of the wetting layer on the curved side decreases with self-propulsion speed above a threshold.
We show that speed diversity amplifies active matter rectification. This is because the contribution from fast particles to rectification increases superlinearly with self-propulsion speed due to two cooperative effects. The first cooperative effect is that the rectification wind pushes particles moving in the `wrong' direction and force them to move along the `easy' rectification direction. Once inside the wetting layer, rotational diffusion decorrelates their self-propulsion orientation and those particles end up escaping along the $x$ axis upon reaching the obstacle corners. The second cooperative effect is that the rectification wind also reduces the amount of particles trapped in the wetting layer on the curved side, contributing to rectification. Our results also complement the explanation given in Ref.~\cite{potiguar2014self} for the rectification of monodisperse systems: particles coming from the left side become rectified when sliding along the curved side of the obstacle, whereas the particles coming from the right side are reoriented by vortices rotating favorably to the global current near the corners. Also, in the passive-active case, we observe that no complete segregation is achieved and that the passive particles continue to be rectified because they are pushed by the active ones.
The persistent rectification current modifies the absorption and evaporation rates on the wetting layers. As a result, the lever rule of equilibrium thermodynamics and MIPS is not satisfied and thus the density of the gas phase increases with the global density. Such violation of the lever rule increases with speed diversity. This is consistent with the idea that the lever rule violation is caused by rectification since the rectification velocity itself also increases with speed diversity.

The present work helps highlight the quantitative importance of considering speed diversity. Usually, self-propulsion speeds are measured but only its average value is used to make new predictions, which can lead to significantly inaccurate values for the rectification velocity. As we show, rectification can get doubled solely by effects of pure speed diversity. Finally, we mention that, due to the various sources of non-linearities, deriving an analytical expression for the rectification of active mixtures is a challenging endeavor. Nevertheless, future work should focus on developing an analytical theory as well as consider other types of interactions, such as alignment and hydrodynamics. A research program like that will improve our understanding about what determines spontaneous rectification in active matter in more realistic scenarios with complex interactions. 

More generally, the results presented here and in other articles, put in evidence the need of building appropriate theories for mixtures of non-equilibrium entities as active particles: the simple use of an average (effective) fluid is not sufficient and large dependences on the diversity are obtained. Hydrodynamic theories of active mixtures should be able to qualitatively predict the amplification of rectification induced by diversity as well as the spatial structure of the velocity and vorticity fields presented in this article.

\backmatter

\bmhead{Supplementary information}
Movie 1: Transient dynamics for $\phi=0.13$ and $\delta=0.8$.
Movie 2: Transient dynamics for $\phi=0.26$ with $\delta=0.8$.
Movie 3: Steady-state dynamics for $\phi=0.26$ with $\delta=0.8$.

\bmhead{Acknowledgments}

M.R.-V.~and R.S.~are supported by Fondecyt Grant No. 1220536 and Millennium Science Initiative Program NCN19\_170D of ANID, Chile. P.d.C.~was supported by Scholarships No.~2021/10139-2 and No.~2022/13872-5 and ICTP-SAIFR Grant No.~2021/14335-0, all granted by S\~ao Paulo Research Foundation (FAPESP), Brazil.

\bmhead{Author contribution statement} 
Original conceptualization: Pablo de Castro; Simulations: Mauricio Rojas-Vega; Formal analysis and investigation: Mauricio Rojas-Vega, Pablo de Castro and Rodrigo Soto; Writing - original draft preparation: Mauricio Rojas-Vega and Pablo de Castro; Writing - review and editing: Mauricio Rojas-Vega, Pablo de Castro and Rodrigo Soto; Supervision: Pablo de Castro and Rodrigo Soto.

\bmhead{Data availability}
The datasets generated during and/or analysed during the current study are available from the corresponding author on reasonable request.

\bmhead{Conflict of interest}
The authors declare no conflict of interest.

%\section*{Declarations}

%Some journals require declarations to be submitted in a standardised format. Please check the Instructions for Authors of the journal to which you are submitting to see if you need to complete this section. If yes, your manuscript must contain the following sections under the heading `Declarations':

%\begin{itemize}
%\item Funding
%\item Conflict of interest/Competing interests (check journal-specific guidelines for which heading to use)
%\item Ethics approval 
%\item Consent to participate
%\item Consent for publication
%\item Availability of data and materials
%\item Code availability 
%\item Authors' contributions
%\end{itemize}
%
%\noindent
%If any of the sections are not relevant to your manuscript, please include the heading and write `Not applicable' for that section. 

%\bibliographystyle{unsrt}

\bibliography{Active.bib}

\end{document}